\documentstyle[prl,twocolumn,aps]{revtex}
\begin{document}
\draft 
\title{\large  \bf
Langevin approach to generate synthetic turbulent flows
}
\author{\large A.C. Mart\'\i ,$^{1}$ J.M. Sancho,$^{1}$
F. Sagu\'es$^{2}$ and A. Careta$^{2}$ }

\address{
$^1$ Departament d'Estructura i Constituents de la Mat\`eria,
Universitat de Barcelona,\\Av. Diagonal 647,
E-08028 Barcelona, Spain.\\
$^2$ Departament de Qu\'{\i}mica F\'{\i}sica,
Universitat de Barcelona,\\ Av. Diagonal 647,
E-08028 Barcelona, Spain.\\
}

\date{\today}

\maketitle 

\begin{abstract}
We present an analytical scheme, easily implemented  numerically, 
to generate synthetic  Gaussian turbulent
flows by using a linear Langevin equation, where the noise term acts as a 
stochastic stirring force. The characteristic parameters of
the velocity field are well introduced, in particular the kinematic viscosity
and the spectrum of energy. 
As an application, the diffusion of a passive scalar is studied for two 
different energy spectra. Numerical
results are compared favorably with analytical calculations.

\end{abstract}

\vskip5mm

\section{Introduction} 
\label{sec:intr}
The statistical approach to turbulence has a long history on its own 
\cite{Monin75,Batchelor70}.
Actually, it is the firm recognition that both fundamental and applied aspects
of turbulence can be consistently addressed from the knowledge of the
statistical moments and correlations of the velocity flow that
has made this approach particularly
appealing. Since in any case such an statistical point of view has 
adopted  multiple perspectives during these last decades, let us first put our 
formulation into an appropriate context.

Our primary aim here is to describe an analytical and numerical 
methodology to generate a statistically stationary, homogeneous, and isotropic 
two-dimensional turbulent flow with zero mean velocity and well--defined  
energy spectrum. Needless to say, such a regime of steady turbulence can only be 
maintained by means of an external input of energy to compensate the 
dissipative nature of the viscous forces.
This external input 
could be conceptually
associated with stirring forces which are stochastic in nature 
so as to produce a random velocity field which is to represent the turbulent 
flow \cite{McComb90and95}. 
It is quite obvious that the statistical properties of the steady stirring 
forces will ultimately determine those of the turbulent flow.

Ideally, the statistical properties of any turbulent flow should 
come as the output of a first principles, Navier-Stokes based, formulation
of the problem. However, we will adopt here a somewhat reversed perspective
aimed at developing a methodology to construct what would be 
a sort of synthetic turbulence \cite{Careta93B,Juneja94,Siggia94,Kaplan88,Pope}.
Rather than  to retain 
the nonlinear coupling which makes possible the redistribution of energy 
from the largest 
length scales down into the smaller ones, we assume that
the energy is incorporated
into the system in an individual wave number basis. 
In other words, our model would 
represent a collection of uncoupled stirrers each one acting on a single--length 
scale  and introducing its own wave number dependent energy 
contribution, but chosen 
to reproduce, and this is
the main point in our approach, the
desired spectral distribution in steady state.

This practical approach to produce a turbulent 
flow is justified by the fact that  what we want is to study a physical 
process within the turbulent medium described only by its
statistical properties.
Our  strategy is accomplished by using a generalized Ornstein Uhlenbeck--like
Langevin equation for the stream function  $\eta ({\bf r},t)$ 
\cite{Careta93B,Siggia94}. 
Being more specific, our proposal is based on the use of the following Langevin equation
\begin{equation}
{\partial \eta({\bf r},t) \over {\partial t}} =
 \nu  \nabla^2 \eta({\bf r},t)
+Q[\lambda^2 \nabla^2] \nabla . \mbox{\boldmath $ \zeta $}  ({\bf r},t),
\label{langevin}
\end{equation}
where $\nu$ is the kinematic viscosity and
$\mbox{\boldmath $ \zeta $} ({\bf r},t)$
is a Gaussian white noise with zero mean value and  correlation

\begin{equation}
\left< \zeta^i({\bf r}_1,t_1)  \zeta^j({\bf r}_2,t_2) \right> = 2
\epsilon_0 \nu  
\delta (t_1-t_2)
\delta ( {\bf r}_1 - {\bf r}_2) \delta^{ij}.
\label{ruido} 
\end{equation}
In this last expression, $\epsilon_0$, and  $\lambda$ are
parameters which control, respectively, the intensity and 
correlation length of the random flow. $Q[\lambda^2 \nabla^2]$ 
will play the most relevant role in our scheme, representing the 
random stirring forces.

The random flow generated in this way  has 
Gaussian properties due to the linear nature of
such an equation. 
Apart from that, other limitations concerning some invariances of the 
turbulent flow appear also as a consequence of such a linearity 
\cite{Siggia94}. However, the big advantage of our model is that it 
facilitates the control over the characteristic parameters of the 
turbulence, 
i.e. its integral time and length scales and its spectrum, just by 
appropriately prescribing the input parameters of the noise entering into 
the Langevin equation. A final remark concerning Gaussianity is worth 
mentioning at this point. Certainly, we recognize that such an 
statistical property is under scrutiny \cite{She91}. However 
particular experimental scenarios supporting it \cite{Ronney95}, together 
with the lacking of conclusive results on intermittency 
\cite{McComb90and95}, leave this question rather open and allows us to use 
such a Gaussian property, at least as a working simplifying hypothesis,
mainly when the focus is in the study of physical 
process inside this medium.

This paper is organized as follows. In the next section we summarize
the Langevin approach to generate the random flow (synthetic turbulence). 
In Sec. \ref{sec:sim}   the
technical details of the numerical simulations are presented. 
As a sort of application  we consider the classical problem of the diffusion
of a passive scalar in Sec. \ref{sec:diff}. We devote Sec.  
\ref{sec:summ} to draw some  conclusions and perspectives.
\section{Langevin Approach}
\label{sec:lang}
As our approach starts with the generation of a scalar stream function 
\cite{Landau87}, we 
are going to review first the relationships between the stream function 
properties 
and those of the velocity field. Let $\eta ({\bf r},t)$ be the stream 
function from which we define the  two--dimensional
incompressible  turbulent field ${\bf 
v}({\bf r},t)$ (${\bf \nabla}.{\bf v}=0$),
\begin{equation}
{\bf v }({\bf r},t) = \left( -{{\partial \eta ({\bf r},t)} \over {\partial y}}
,{{ \partial \eta ({\bf r},t)} \over{ \partial x }} \right) .
\label{defvel}
\end{equation}
Its mean value is to be taken zero and the two-point velocity 
autocorrelation is defined as usual through

\begin{equation}
\left< v^i ({\bf r}_1,t_1)  v^j ({\bf r}_2,t_2) \right> =
R^{\ ij}({\bf r}_1-{\bf r}_2,t_1-t_2),
\end{equation}
where homogeneity in space and time is explicitly denoted 
\cite{Batchelor70,McComb90and95}.
Our scheme is entirely implemented in two dimensions (2D). However,
it could be    generalized to 3D just by taking a vector
stream function  
$\mbox{\boldmath $ \eta $}$
and obtaining the velocity field
as $  {\bf v}= \nabla \times
\mbox{\boldmath $ \eta $}$.
In this case each component
of the stream function must satisfy a Langevin equation
similar  to (\ref{langevin})  with independent noises.

Making use of the spatial isotropy we define the radial 
correlation function as 
\begin{equation}
R(r,s) =  {1 \over{2}} [R^{\ xx}(r,s) +R^{\ yy}(r,s)],
\label{correlationradial}
\end{equation}
where  $r=|{\bf r}_1 - {\bf r}_2| $ and $s=|t_1 - t_2|$.
The autocorrelation of $\eta ({\bf r},t)$  
\begin{equation}
 C(r,s) = \left< \eta({\bf r}_1,t_1) \eta({\bf r}_2,t_2) \right>
\label{correlationstream}
\end{equation}
is correspondingly assumed  to have the properties 
of homogeneity, isotropy,  and 
stationarity. From the definition (\ref{defvel}) we can express 
the velocity correlation (\ref{correlationradial}) in terms of 
(\ref{correlationstream})
\begin{equation}
R(r,s) = \frac{1}{4 \pi} \int_{0}^{\infty } {\! dk \, k^3 J_0(kr) 
\mathaccent22{C}(k,s)},
\label{correlacioncalculada}
\end{equation}
where $J_0(kr)$ is the Bessel function of zeroth order and
$\mathaccent22{C}(k,s)$ is the Fourier transform of  $C(r,s)$.

The physical parameters of steady turbulence, i.e.,
its intensity $u_0^{\ 2}$, 
and characteristic (integral) time and length scales follow from
their standard definitions \cite{Batchelor70,McComb90and95}, 

\begin{eqnarray}
u_0^{\ 2}    & =   &  R(0,0), \nonumber \\ 
t_0 &  = &  {1 \over { u_0^{\ 2}}} \int_0^{\infty } ds \  R(0,s),\\
\label{physpar}
l_0 & = & {1 \over { u_0^{\ 2}}} \int_0^{\infty } dr \ R(r,0).
\nonumber
\end{eqnarray}

With these definitions in mind let us move to the discussion of the 
analytical scheme. 
The Fourier transform of (\ref{langevin}) reads 
\begin{equation}                                        
{\partial \eta({\bf k},t) \over {\partial t}} = - \nu  k^2 \eta({\bf k},t)
-i Q(- \lambda^2 k^2) k^j  \zeta^j({\bf k},t).
\label{langevinfourier}
\end{equation}
Now we introduce the structure function $S(k,t)$  related with
the equal--time 
correlation 
function (\ref{correlationstream}), but in Fourier space
\begin{equation}
\left< \eta({\bf k}_1,t) \eta({\bf k}_2,t) \right> =
(2 \pi)^2 \delta ({\bf k}_1 + {\bf k}_2) S(k,t).
\label{correlFour}
\end{equation}
By using standard stochastic calculus \cite{Gardiner85},  $S(k,t)$ 
is shown to verify
\begin{equation}
{\partial S(k,t) \over{ \partial t }}= -2 \nu k^2 S(k,t)  +
2 \epsilon_0 \nu k^2 Q^2 (- \lambda^2 k^2).
\label{S(k)}
\end{equation}
On the other hand, $\mathaccent22{C}(k,s)$ obeys, 
in the steady state, the 
equation 
\begin{equation}
{\partial \mathaccent22{C}(k,s) \over{ \partial s }}= - \nu k^2 
\mathaccent22{C}(k,s),
\label{C(k,s)}
\end{equation}
with the initial condition 

\begin{equation}
\mathaccent22{C}(k,0) =  S(k,t \to \infty) = 
S_{st}(k) \label{C(k,0)}.
\end{equation}
From (\ref{S(k)}) and (\ref{C(k,s)}) we get 
\begin{equation}
\mathaccent22{C}(k,s) = \epsilon_0  Q^2 (- \lambda^2 k^2) e^{- \nu k^2 s},
\label{Ccalc}
\end{equation}

The energy spectrum $E(k,t)$ is defined in terms of $S(k,t)$ as       
\begin{equation}
E(k,t) = {1 \over {4 \pi}} k^3 S(k,t) .
\label{EyS}
\end{equation}
Using (\ref{S(k)}) we can also obtain the equation 
of evolution  of $E(k,t)$ \cite{Batchelor70,McComb90and95,Breuer94}
\begin{equation}
{d E(k,t) \over {d t}} = -2 \nu k^2  E(k,t) + 2 k^2 W(k),
\end{equation}
where
\begin{equation}
W(k)= { \epsilon_0 \nu \over {4 \pi}} k^3 Q^2 (- \lambda^2 k^2).
\end{equation}
This quantity can be regarded as 
the input of energy due to the stirring forces. The stationary state  
is achieved when  the input term is exactly balanced by the dissipation term,
\begin{equation}
E(k,t \to \infty) = E(k) = {1  \over {\nu}} W(k)=
{ \epsilon_0 \over {4\pi}}k^3 Q^2 (- 
\lambda^2 k^2) = {1  \over {4\pi}} k^3 \mathaccent22{C}(k,0).
\end{equation}
The stirring intensity $u_0^2$ can also be related with
the spectrum
\begin{equation}
u_0^{\ 2} =  \int_0^{\infty }{ dk \ E(k)},
\end{equation}
and the characteristic time $t_0$ (\ref{physpar}), and the  
characteristic length  (\ref{physpar})
can be related as well with  the viscosity $\nu$  and
$\lambda$. Explicit relations will be obtained later on for particular 
spectra.

As it is quite obvious, our scheme does not incorporate the non--linear term
of the Navier-Stokes equation. Notice, nevertheless, that this scheme is
versatile enough to reproduce a large variety of energy spectra, just by
appropriately prescribing the differential operator $Q[\lambda^2 \nabla^2]$
in (\ref{langevin}). In our approach the noisy term represents the stirring
mechanism which feeds energy continuously into the system according to the
chosen spectrum. At the same time this energy is dissipated at time and
spatial scales controlled by $\nu$. As there are only linear terms in the
Langevin equation, no kind of energy cascade is possible. The
noise forces introduce the 
whole hierarchy of turbulent structures evolving according to  their own 
time and length scales, but without any interaction between them. 

\section{Simulation Algorithm}
\label{sec:sim}

We have chosen for the discretization of the real space 
a standard two--dimensional square lattice $N \times N$
with elementary unit spacing $\Delta$ in both directions. In most of our
 simulations $N=128$ and  $\Delta = 0.5$, unless other values are specified.
 The discrete 
Fourier space is discretized accordingly  ${\bf k} = (k_x , k_y ) =
2 \pi /  N \Delta  ( \mu , \upsilon ) $ \cite{Ojalvo92} 
(Greek indices are used in Fourier space on all that follows).
We can take advantage of the fact that (\ref{langevinfourier}), when 
written in the discrete Fourier space, transforms  into a 
set of linear and not
coupled ordinary differential equations. In these circumstances the exact
integration between $t$ and $t + \Delta t$, $( \Delta t = 0.1 t_0 )$ gives 
\begin{equation}
\eta_{\mu \upsilon}(t+ \Delta t) =   \exp ( \nu c_{\mu \upsilon} \Delta t ) \,
\eta_{\mu \upsilon}(t)
+ \beta_{\mu \upsilon}(t)+ \gamma_{\mu \upsilon}(t),
\end{equation}
in terms of new random variables $\beta_{\mu \upsilon}(t)$
 and  $\gamma_{\mu \upsilon}(t)$ are defined according to
\begin{equation}
\beta_{\mu \upsilon}(t)=
Q_{\mu \upsilon} d_{\mu \upsilon}^x \int_t^{t+\Delta t} {dt' \, \zeta_{\mu
 \upsilon}^x (t') \exp\left[(t+\Delta t -t') \nu c_{\mu \upsilon} \right]},
\end {equation}
 \begin{equation}
\gamma_{\mu \upsilon}(t)=
Q_{\mu \upsilon} d_{\mu \upsilon}^y 
\int_t^{t+\Delta t}
{dt' \, \zeta_{\mu \upsilon}^y (t') \exp\left[(t+\Delta t -t')
\nu c_{\mu \upsilon} \right]},
\end {equation}
where $Q_{\mu \upsilon}$ in the last two expressions denote the 
discrete Fourier 
transform of the operator $Q[\lambda^2 \nabla^2]$.
In the Fourier space the derivative operators have been translated into
\begin{equation}
\nabla^2 \rightarrow c_{\mu \upsilon} = {2 \over {\Delta^2}}
\left[ \cos \left({2 \pi \mu \over {N}}\right) +
\cos \left({2 \pi \upsilon \over {N}}\right) -2 \right],
\end{equation}

\begin{equation}
\nabla .  \rightarrow {\bf d}_{\mu \upsilon} = {1 \over {\Delta}}
\left( \exp \left( {\rm i} { 2 \pi \mu \over {N}}\right) -1, 
 \exp  \left( {\rm i}  {2 \pi \upsilon \over {N}}\right) -1
\right).
\end{equation}

The correlation of the random variables can be expressed as 
\begin{eqnarray}
\left< \beta_{\mu \upsilon}^*(t) \beta_{\rho \sigma}(t) \right> = 
Q_{\mu \upsilon} Q_{\rho \sigma} d_{\mu \upsilon}^x d_{\rho \sigma}^{x*}
< \zeta_{\mu \upsilon}^{x*} (t') \zeta_{ \rho \sigma}^{x} (t'') >
\times 
\nonumber
\\
\int_0^{t+\Delta t}{dt'  \int_0^{t+\Delta t}{dt'' 
\exp \left[(t+\Delta t -t') \nu c_{\mu \upsilon}+ \nu c_{\rho \sigma}
(t+\Delta t -t'' ) \right]}},
\end{eqnarray}
and similar equations for the $\gamma_{\mu \upsilon}(t)$.
Using (\ref{ruido}), the symmetry properties of the $Q$ operator $ Q_{\mu 
\upsilon} = Q_{- \mu - \upsilon}= Q_{\mu \upsilon}^* $,  
and further integrating and using the expressions defined above for 
the discrete operators in 
Fourier space we finally have
\begin{equation}
\left< \beta_{\mu \upsilon}^*(t) \beta_{\rho \sigma}(t) \right> =
2 \epsilon_0  
\delta_{\mu \rho } \delta_{\upsilon \sigma} N^2
Q_{\mu \upsilon}^2  c_{\mu \upsilon}^{-1}
\left[  \exp \left( 2  \nu c_{\mu \upsilon} \Delta t  \right) -1 \right]
\left[ 1- \cos\left({2 \pi \mu \over{N}}\right) \right].
\end{equation}
We can now construct an explicit expression for $ \beta_{\mu \upsilon}(t) $
 adapted to the result just obtained
\begin{equation}
\beta_{\mu \upsilon}(t)  =
\left( 2 \epsilon_0  N^2
Q_{\mu \upsilon}^2  c_{\mu \upsilon}^{-1}
\left[  \exp \left( 2  \nu c_{\mu \upsilon} \Delta t  \right) -1 \right]
\left[ 1- \cos\left({2 \pi \mu \over{N}}\right) \right] \right)^{1/2}
\alpha_{\mu \upsilon}(t),
\end{equation}
where $\alpha_{\mu \upsilon}$ are Gaussian random numbers, which satisfy
$\left< \alpha_{\mu \upsilon}^*(t) \alpha_{\rho \sigma}(t) \right> =
\delta_{\mu \rho } \delta_{\upsilon \sigma}$.
We would proceed analogously for  $ \gamma_{\mu \upsilon}(t)$.
The correlation of the  stream function in the steady state is
\begin{equation}
\left< \eta_{\mu \upsilon}^* \eta_{\rho \sigma} \right>_{st} =
Q_{\mu \upsilon}^2 \epsilon_0 (N \Delta)^2 
 \delta_{\mu \rho } \delta_{\upsilon \sigma}.
 \label{larget}
\end{equation}
When dealing with the diffusion of the passive scalars in Sec.
\ref{sec:diff} we 
will always  start our simulations in a steady configuration of the 
random flow. According to the result (\ref{larget}) this is easily 
accomplished by taking as initial condition

\begin{equation}
\eta_{\mu \upsilon}(0)  =
Q_{\mu \upsilon} \epsilon_0^{1/2} (N \Delta) \alpha_{\mu \upsilon}(0).
\label{steadystate}
\end{equation}

At each time step and after having generated the 
 stream function in Fourier space we proceed to antitransform 
and using  it in relation with the appropriate discretized forms for
 the velocity field.
We skip some details to refer directly to the 
 the discrete version of the 
energy spectrum which finally reads  \cite{Careta93B}
\begin{equation}
E_{\mu \upsilon} = {\epsilon_0 \over{ 2 N  \Delta^3}} 
( \mu^2 + \upsilon^2)^{{1\over{2}}} 
Q_{\mu \upsilon}^2
\left[ \sin ^2
\left( {2 \pi \mu \over{ N }} \right) + \sin^2 \left( {2 \pi \upsilon
\over{ N }} \right) \right].
\end{equation}

\section{Diffusion of passive scalars}
\label{sec:diff}

Before we start the study of the scalar diffusion, we will discuss
the selection of spectra. There are several possible choices
and between all of them we have selected two spectra which are well behaved
in all the range of wave numbers.

\subsection{Kraichnan's and K\'arman-Obukhov spectra}
The spectrum introduced by Kraichnan (K) \cite{Kraichnan70} describes 
turbulent velocity fields  with a 
widely distributed band of excitations and a  peak centered at well-defined 
wave number $\sim k_0$
\begin{equation}
E(k) \propto k^3 \exp \left[ - { k^2 \over{ k_0^2}} \right].
\label{Kraichnan}
\end{equation}
In this case, the operator $Q$ has to be chosen according to                               
\begin{equation}
Q[ \lambda^2 \nabla^2] = \exp \left( - {\lambda^2 \nabla^2 \over {2}} \right),
\end{equation}
where $\lambda = k_0^{-1}$.
Using (\ref{Ccalc}) and substituting in (\ref{correlacioncalculada}),
 the velocity correlation function in steady state reads
\begin{eqnarray}
R(r,s) = \frac{\epsilon_0}{8 \pi  (\lambda^2 + \nu s)^2}
\left[1 - {r^2 \over{4 (\lambda^2 + \nu s)}} \right] \exp
\left[ - {r^2 \over{4 (\lambda^2 + \nu s)}} \right].
\end{eqnarray}
From it, we can obtain  $u_0^2$, 
and the integral time and length scales according to
\begin{equation}
u_0^2 = \frac{\epsilon_0}{8 \pi \lambda^4},
\end{equation}
\begin{equation}
t_0 =  \frac{\lambda^2}{\nu},
\end{equation}
\begin{equation}
l_0 =  \frac{\lambda \sqrt{ \pi }}{2}.
\end{equation}
The dimensionless  Reynolds number, defined according to 
\begin{math}
{\rm Re} = l_0 u_0 / \nu,
\end{math}
 is thus expressed in terms of the noise parameters as
\begin{equation}
{\rm Re}= \left( {\epsilon_0 \over{2}} \right)^{1/2} { 1 \over{4 \lambda \nu}}.
\end{equation}

Our second choice is the K\'arman-Obukhov's (KO) spectrum which was
introduced \cite{K-O} 
to study Kolgomorov turbulence
with a long $``-5/3''$ tail in the spectrum for large $k$. 
To this end we select the family of K\'arman-Obukhov spectra, whose general
behavior is
\begin{equation}
E(k) \propto k^n \left[ 1 + { k^2 \over{ k_0^2}} \right] ^{
-(5+3n)/6}.
\label{eq:K-O}
\end{equation}
For convenience we have taken $n=3$.
The choice of the  $Q$ operator is then  

\begin{equation}
Q[ \lambda^2 \nabla^2] =  \left( 1 - \lambda^2 \nabla^2  \right)^{-7/6}.
\label{eq:QforK-O}
\end{equation}
In this case we obtain the following results for the three basic parameters
\begin{equation}
u_0^2 = \frac{9 \epsilon_0}{32 \pi \lambda^4},
\end{equation}
\begin{equation}
t_0 =  \frac{\lambda^2}{3 \nu},
\end{equation}
\begin{equation}
l_0 = \lambda {\Gamma(1/2) \Gamma(5/6) \over{2 \Gamma(1/3)}}.
\end{equation}
The corresponding Reynolds number is expressed as  
\begin{equation}
{\rm Re }= \left( {\epsilon_0 \over{2}} \right)^{1/2} { 1 \over{4 \lambda \nu}}
{3 \Gamma(5/6) \over{2 \Gamma(1/3)}}.
\end{equation}
A closed expression for the correlation function cannot be obtained
in this case.

These two spectra have been simulated according to the recipes of the
previous section. In Fig.~1 we have plotted 
the evolution of an initially flat
spectrum towards its final steady state. By looking at the patterns 
corresponding to different time intervals
we can judge the role of the kinematic viscosity.
In particular the modes with smaller $k$ relax slower than modes with
larger $k$.
Continuous and discrete analytical results are also compared in the
steady state to see the differences introduced by the discretization
procedure specially for large $k$.
The clear differences between both spectra are going to influence the 
diffusion  of passive scalar as it will be discussed later on.

\subsection{ Scalar diffusion in a continuous scheme}

Scalar diffusion in turbulent flows is a classical problem 
\cite{Taylor21} which still deserves a lot of attention \cite{Diffusion}.
Our starting point is the Eulerian equation of motion for  a
scalar distribution  $\psi ({\bf r},t)$ which is assumed to be passively 
advected by the previously prescribed isotropic homogeneous  and 
stationary random flow ${\bf v} ( {\bf r} , t )$,

\begin{equation}
{\partial \psi ({\bf r},t) \over{ \partial t}}= 
D \nabla^2 \psi ({\bf r},t) - {\bf \nabla}.[{\bf v}({\bf r},t)
\psi ({\bf r},t)],
\label{Dev}
\end{equation}
In the standard notation used in (\ref{Dev}), 
$D$ stands for the ``bare'' 
molecular
diffusion coefficient. 
By averaging over 
realizations of ${\bf v}({\bf r},t)$ we simply obtain from the above equation 
the temporal evolution of $<\psi ({\bf r},t)>$. Actually taking a 
$\delta$-like initial condition, this quantity is nothing but the 
probability density for the spatiotemporal dispersion of a unit amount 
of the randomly advected scalar \cite{Careta93A,Careta94}. Thus the first 
nonzero moment,
\begin{equation}
<r^i r^j>_{<\psi ({\bf r},t)>}=
\int_{R^n}{dv  \, r^i r^j <\psi ({\bf r},t)>},
\end{equation}
is all that we need to compute  an effective diffusion coefficient.
The procedure outlined above, although simply enunciated, is quite 
involved in its analytical resolution. The best way to proceed is thus to 
follow the standard analytical strategies furnished by the theory of 
Gaussian stochastic processes. The central difficulty arises from the 
non-Markovian nature of the process at hand. Controlled perturbative 
schemes  are thus necessary. In particular a consistent 
expansion, \cite{Careta94} based on the smallness of the correlation time, 
$t_0$, leads to a closed equation for $<\psi ({\bf r},t)>$, linear in the 
autocorrelation tensor, from which a diffusive regime is identified 
through the common linear law for the scalar dispersion

\begin{equation}
 <r^2> - <r>^2 = <(\Delta r)^2> = 4 D_{ \rm eff} t = 4 (D +  \Delta D) t.
\end{equation}
Moreover from such an expansion, the explicit expressions for the leading 
contribution to $\Delta D$ can be simply evaluated as 
\begin{equation}
\Delta D_c = \int_0^{\infty}{ds \, R(0,s)} + 
4 D \int_0^{\infty}{ds \, s  R''(0,s)},
\label{DeltaDc}
\end{equation}
where $ R''(0,s) = \partial ^2 R(r,s) / \partial r^2 \vert_{r=0}$.
Computing these left integrals for both spectra we end up with a common
expression which reads
\begin{equation}
\Delta D_{c} = u^2_0 t_0 
\left(1- {2 D \over{\nu}} \right).
\end{equation}
In particular for the Kraichnan's flow
\begin{equation}
\Delta D_{c K} =u_0^2 t_0 \left(1-{D \pi t_0 \over{2 l_0^2}}
\right).
\label{eq:DeltaDcK}
\end{equation}
An analogous expression would be obtained for the K\'arman-Obukhov's 
spectrum.

Both expressions identify the zeroth--order contribution,
$u_0^2 t_0$, which can be also viewed as an exactly  correct limiting 
case of the classical Roberts analysis \cite{Roberts61} for the diffusion 
of a scalar field advected by a rapidly varying random velocity field.

\subsection{Discrete scheme}
Needless to say that given the discrete nature of our simulations, the 
numerical results for the scalar dispersion will be more favorably compared
when referring to the discrete version of the analytical results given above. 
In particular (\ref{DeltaDc}) transforms into 
\begin{displaymath}
\Delta D_d = \int_0^{\infty}{ds \,R(\Delta,s)} +
{2 D \over{\Delta^2}} \int_0^{\infty}{ds \,s \left[ R(2^{1/2} \Delta,s) +
R( 2 \Delta,s)
-2 R(\Delta,s) \right] },
\end{displaymath}
For the spectra here analyzed the explicit expression reads  
\begin{eqnarray}
\Delta D_d =  {\epsilon_0 \over{ 2 \nu N^2 \Delta^4}}
\sum_{\mu \upsilon} \left[ \sin^2
\left( {2 \pi \mu \over{ N }} \right) + \sin^2 \left( {2 \pi \upsilon
\over{ N }} \right) \right] {Q^2_{\mu \upsilon} \over{-c_{\mu \upsilon}}}
\nonumber \\
\times
\left[ \exp \left( -  i
{2 \pi \mu \over{N}}\right)+ \exp \left( - i
{2 \pi \upsilon \over{N}}\right) \right]
\nonumber \\
\times
\left[ {1 \over{2}} - {2  D \over{ \Delta^2 c_{\mu \upsilon} \nu}}
\left[  {- 1 \over{2}}+ \exp \left( -  i
{2 \pi \mu      \over{N}}\right)+ \exp \left( - i
{2 \pi \upsilon \over{N}}\right)
\right]   \right].
\label{DeltaDdisc}
\end{eqnarray}

\subsection{Results}
\label{subsec:results}

Numerical simulation proceeds first by constructing the random velocity
field with the desired statistical properties and then by seeding the
system with a $\delta$-like initial condition for the dispersed scalar 
at the center of the lattice.
Equation (\ref{Dev}) is numerically simulated by using a first--order
Euler algorithm for the time variable and
symmetric forms for the derivative 
operators. The values of $\Delta$ and $\Delta t$ where those of Sec.
\ref{sec:sim} which satisfy numeric stability criteria.

As anticipated in Sec. \ref{sec:sim}, the initial condition $\eta ( {\bf 
r},0)$ is chosen to correspond to the steady state of $\eta ( {\bf r},t)$
(understood here in a statistical sense) (see Eq. (\ref{steadystate})). In this
way we are sure to be in an isotropic and homogeneous turbulent
environment from the beginning of the simulation. Randomly advected by the
turbulent velocity field the scalar spreads over the lattice, Fig.~2. At
each time step we measure the variance $<( \Delta r)^2>$ and we fit, after
transients, its temporal evolution to a linear law, to obtain $D_{ \rm eff}$.

Summarized in Fig.~3 we present two series of results for $\Delta D$ 
corresponding to two different choices for the correlation time of the 
random flow $t_0$. Simulation results are compared with the appropriate, 
perturbative obtained theoretical predictions both in their continuous 
and discrete versions. As evidenced in that figure  the agreement is 
entirely satisfactory. The general conclusion is that a large enough 
relative correction to molecular diffusion, up to 25\% in Fig.~3, can be 
accurately predicted as long as the correlation time of the random flow, 
$t_0$, is small enough relative  to the typical time scale for  the scalar 
advection $l_0 / u_0$. 
In addition we note that the differences between the discrete 
and continuous theoretical results are easily  interpreted when comparing 
the corresponding expressions for $\Delta D$. Actually  the correlation 
$R(r,s)$ evaluated at the origin $r=0$ or to first neighbors may be 
significantly different given the small value of $\lambda$ here employed 
($\lambda=1$) relative to the elementary spacing $\Delta$ ($\Delta=0.5$) 
chosen in all our simulations.

The role of the parameter $t_0$ is better analyzed in relation with the 
results of  Fig.~4. Plotted against the inverse of the value of the 
viscosity, a direct measure of $t_0$ for $l_0$ fixed as prescribed here, 
the effective scalar dispersion increases with $\nu^{-1}$. As expected 
from the intrinsic limitation of the perturbative approach here 
developed, numerical simulations and theoretical predictions, although 
showing similar trends, differ progressively as $t_0$ increases. 
Complementary to the results depicted in Fig.~4, chosen to consider the 
role of correlation time  of the random 
flow, we propose Fig.~5 to examine the influence of the correlation 
length $l_0$.
Our theoretical result Eq. (4.19) predicts a very small dependence on 
$l_0$, which is the behavior observed in Fig.~5.
Note in this respect 
that in going from $l_0=1.0 $ to $l_0=3.0$,  $\epsilon_0$ had to be increased 
nearly  
two orders of magnitude to keep $u_0^2$ and  $t_0$
fixed, but  $\Delta D$ hardly changes. 

The last worthy remark refers to the comparison of K and 
KO spectra. To this end we proposed in Fig.~6  a standard 
representation of $\Delta D$ vs. $u_0^2 t_0$ for both flows with 
identical integral values $l_0$ and $t_0$. What we see is that the 
KO spectrum has more statistical dispersion and lower values of the 
effective diffusion than K spectrum. These facts can be understood by 
comparing both spectra in Fig.~1. The most remarkable difference
 between both spectra concerns the inertial subrange they mean to 
represent: KO spectrum is much broader than  K
wavenumber dispersion. This in turn is directly reflected  in the 
different behavior of both spectra for large $k$. 
Thus KO spectrum shows a richer variety  
of turbulent structures at short distances.  For large $k$, KO spectrum
still allows structures of the same order than the lattice size $\Delta$. 
This is not the case of K spectrum where very small structures have a
very low weight as can be seen in the value of the intensity of both
spectra for $k=2$. So in KO spectrum there is less energy concentrated in 
the interval of those $k's$ where the maximum of the energy takes place.
This would finally result in  a reduction of the
effectiveness of  stirring in the case of KO spectrum,
since small $k$-eddies 
can be thought to be more effective in dispersing the scalar.

\section{Summary and Perspectives}
\label{sec:summ}

An stochastic method, based in a Langevin equation, to generate
Gaussian synthetic turbulent flows has been presented.
Characteristic parameters of the flow such as its intensity and integral
time and space scales are well controlled. A relevant aspect of our method is
that multiple choices of flow spectra can be generated. Two of those spectra
are explicitly presented here.
As a practical application of this methodology we have considered
the study of the diffusion of a passive scalar under the influence of 
two different flow spectra.
The role of the
different parameters, in particular the kinematic viscosity,
which control the characteristic of the flow  is discussed. 

The generation of well--controlled flow spectra can be a very useful tool 
in other problems of practical application. For example, this approach was 
already used in the study of phase separation dynamics under
stirring \cite{lacasta}.
Reactive fronts, i.e., flames propagating under turbulent 
convection is presently under study following also this technique \cite{marti}.

\begin{acknowledgements}
We acknowledge financial support by Comision Interministerial de Ciencia y
Tecnologia, (Projects, PB93-0769, PB93-0759) and Centre de Supercomputaci\'o
de Catalunya, Comissionat per Universitats i Recerca de la Generalitat
de Catalunya. A.C.M. also acknowledges partial support from  the CONICYT
(Uruguay) and the Programa Mutis (ICI, Spain).
\end{acknowledgements}

\begin{figure}
\caption{ Dynamical evolution of $E(k,t)$ (dashed lines),
from an initially flat spectrum 
to its steady state;
$t_1= 0.5 t_0$ and $t_2=  t_0$;
K: Kraichnan's, K-O: K\'arman-Obukhov. Solid lines 
correspond to the continuous 
steady  spectra, Eqs. (4.1) and (4.8), respectively, and dotted lines 
correspond to discrete Eq. (3.11)
($u_0^{\, 2} =2.5$, $t_0 = 0.1$, $l_0=1.5$, $100$ realizations). }
\label{fig1}
\end{figure}

\begin{figure}
\caption{ Pattern of an initial black drop of the dispersed
scalar under the influence of
the Kraichnan's flow ($D=0.10$, $\epsilon_0 =651.8$,
$\lambda = 2.26$, $\nu= 5.1$,
and  $ t = 160 $).}
\label{fig2}
\end{figure}

\begin{figure}
\caption{ $\Delta D$ vs  $u_0^{\ 2} t_0$ (Kraichnan's spectrum). 
Here and in the following plots, 
 solid lines correspond to Eq. (4.19), broken lines
to the discrete expression,  Eq. (4.20), and symbols stand
for simulations. Parameter
values: $D =0.30$, $10$ realizations.
Circles: $\lambda =1.00$,  $\nu=5.0$.
Squares: $\lambda =1.69$,  $\nu=28.5$.}
\label{fig3}
\end{figure}

\begin{figure}
\caption{ $\Delta D$ vs the inverse of the kinematic viscosity.
($u_0^2 = 0.25$, $\lambda = 2.0$, $D =0.30$,
 $10$ realizations).}
\label{fig4}
\end{figure}

\begin{figure}
\caption{$\Delta D$ vs  $l_0$. 
($u_0^2 = 0.25$, $t_0=0.10$, $D =0.30$,
$\Delta =0.5 $ 
for the open symbols and short dashed line, $\Delta =1.0$ for the full 
symbols and long dashed line).} 
\label{fig5}
\end{figure}

\begin{figure}
\caption{ $\Delta D$ for
Kraichnan (circles) and  K\'arman-Obukhov (squares) spectra ($t_0=0.10$, 
$l_0=1.5$, 5 realizations).} 
\label{fig6}
\end{figure}

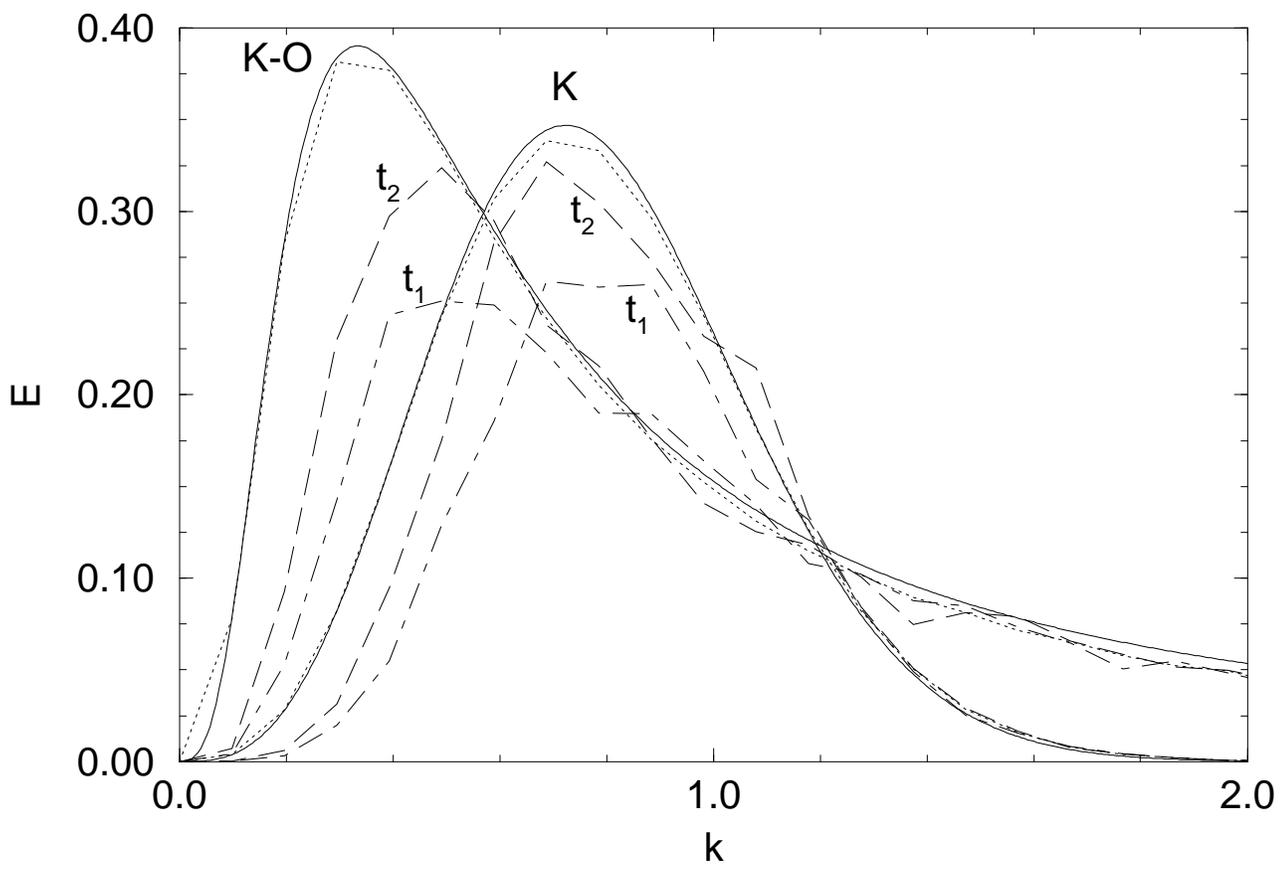
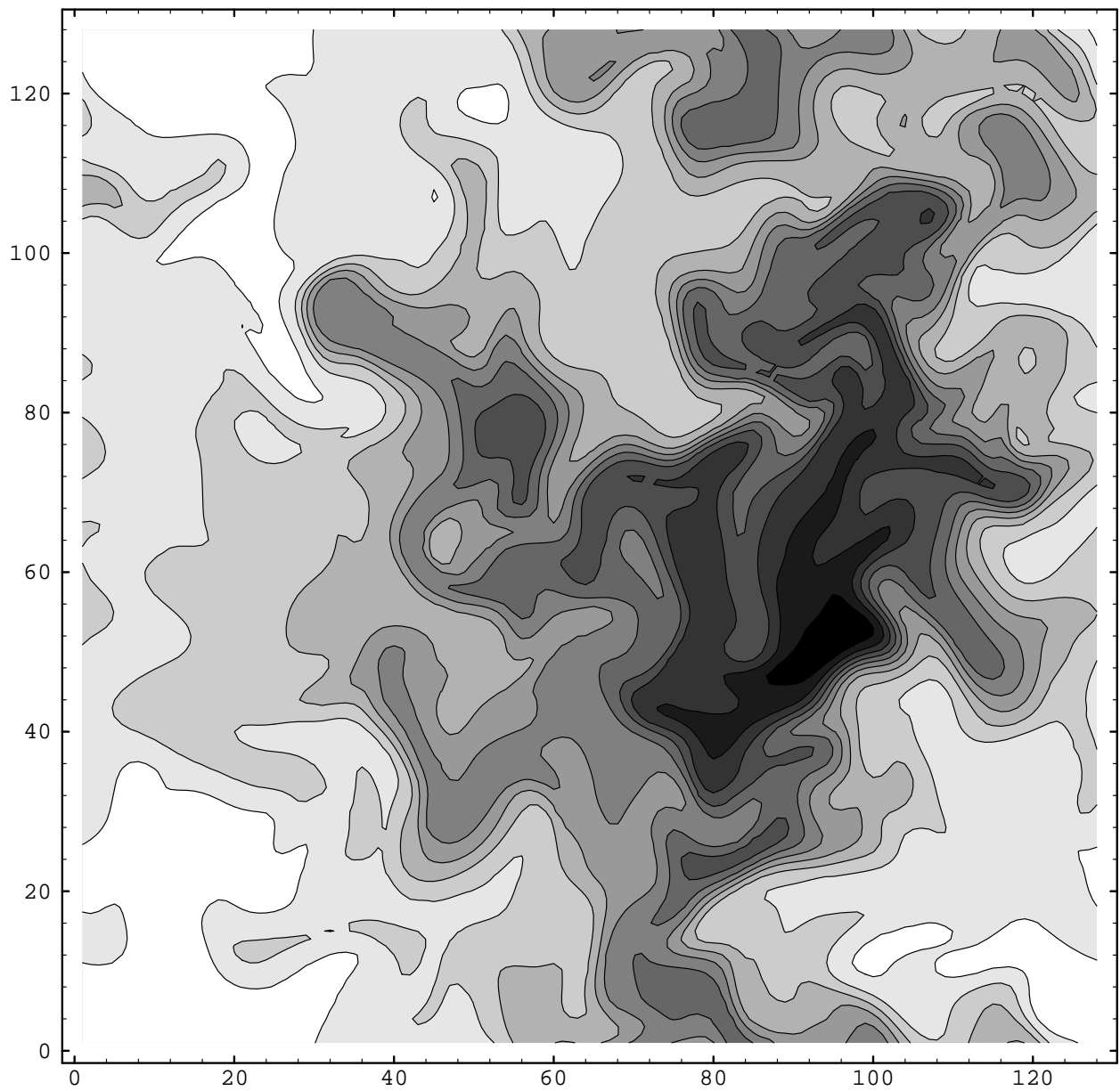
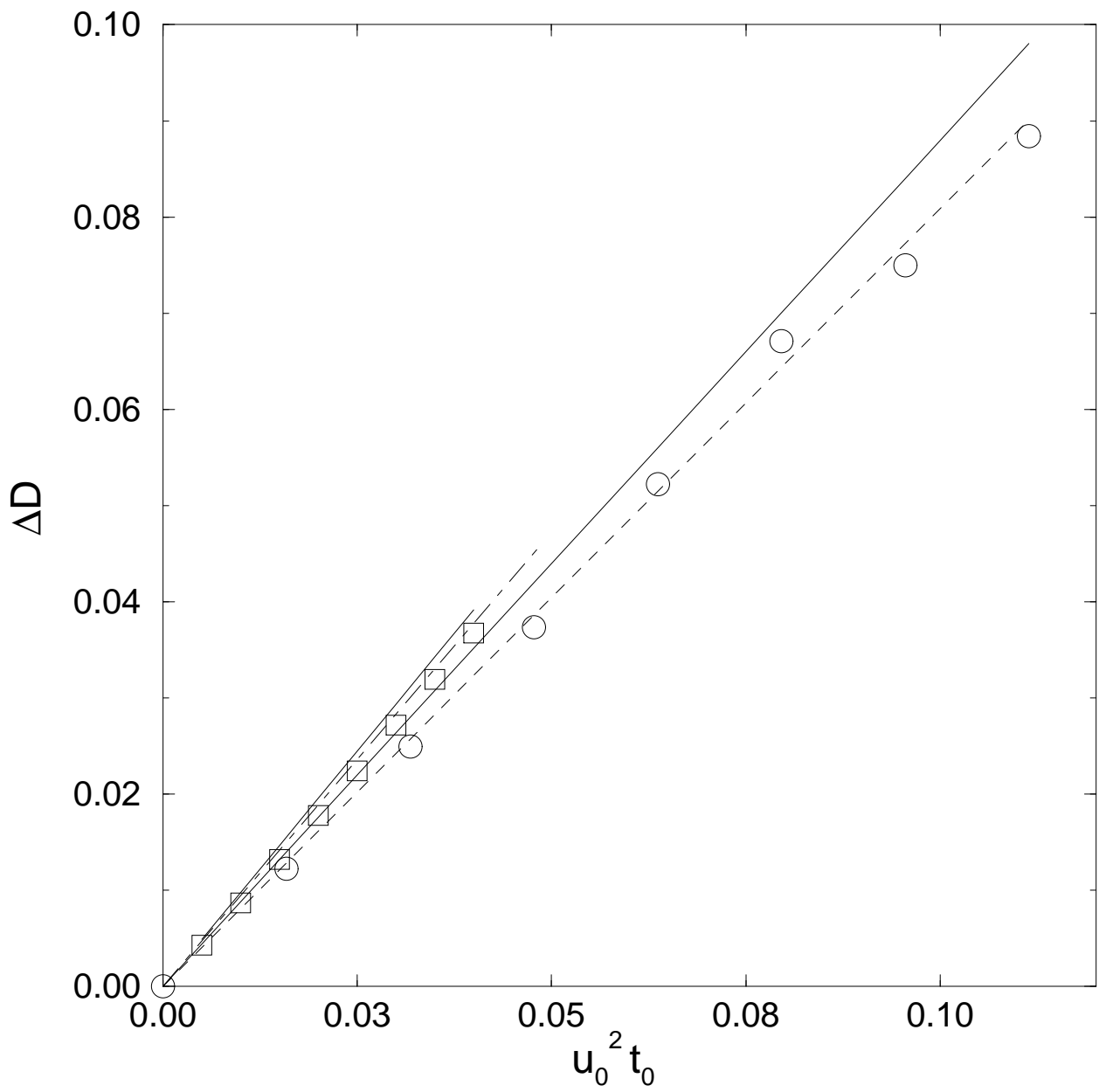
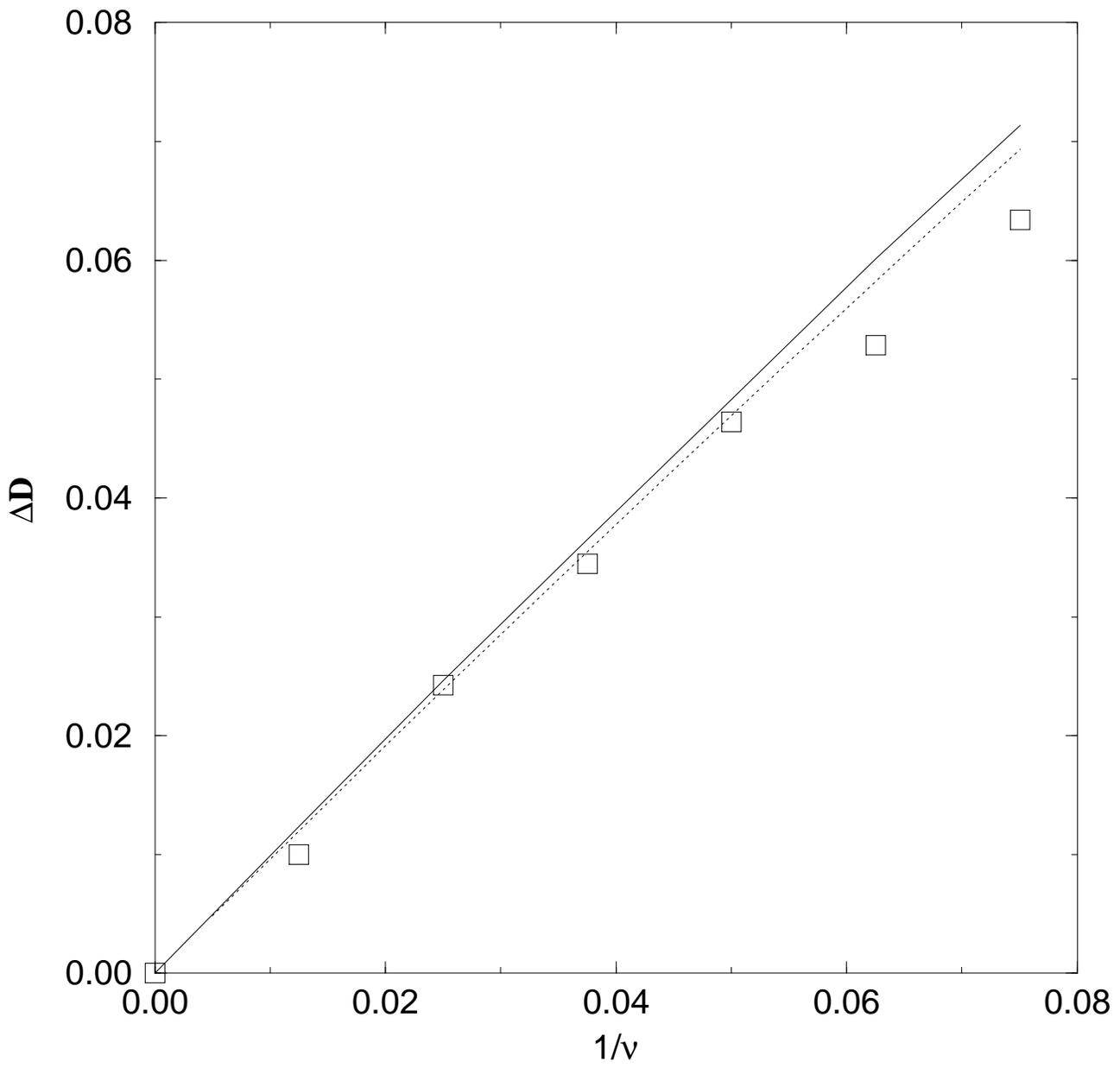
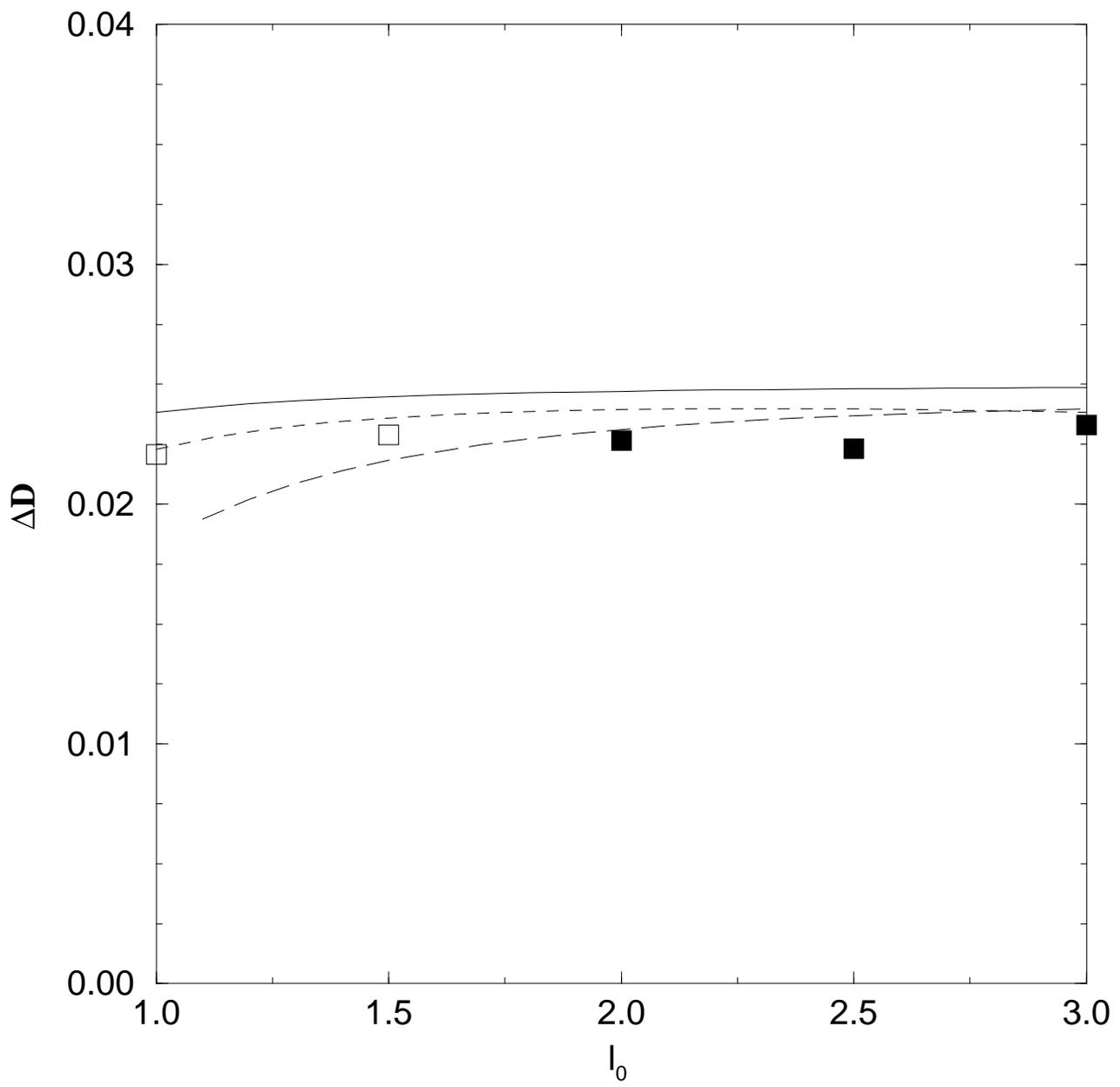
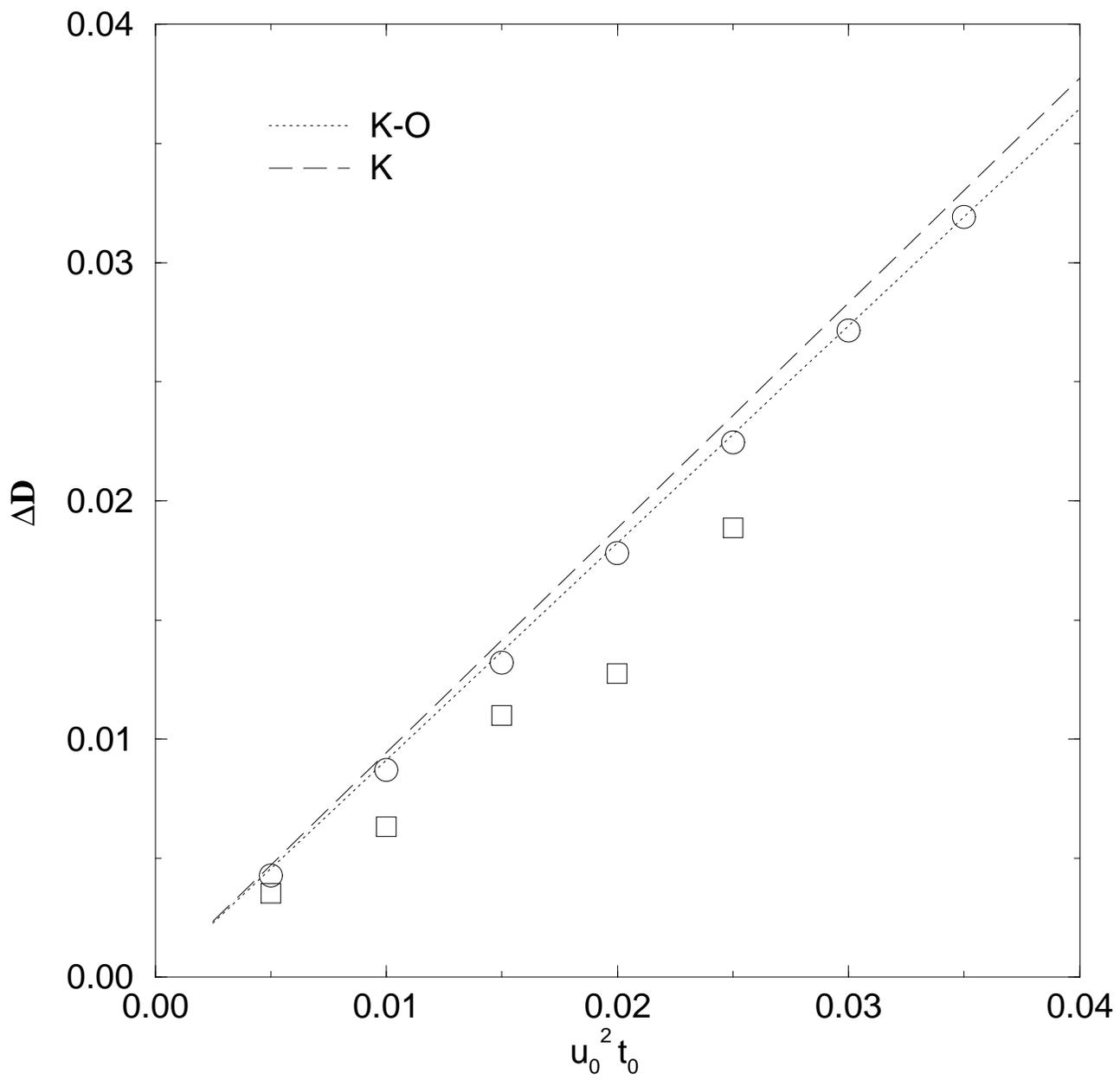
\end{document}